\documentclass[journal]{IEEEtran}
\usepackage{graphicx,amsmath,amssymb,array,balance}
\usepackage{soul} 
\graphicspath{{Figures/}}
\usepackage[numbers,compress]{natbib}

\usepackage{caption}
\usepackage{titlesec}
\titlespacing*{\subsection}{2.5pt}{5pt}{2.5pt}
\titlespacing*{\section}{2.5pt}{5pt}{2.5pt}

\begin{document}

\title{A full-stack view of probabilistic computing with p-bits: devices, architectures and algorithms}

\author{Shuvro Chowdhury, Andrea Grimaldi,  Navid Anjum Aadit, Shaila Niazi,  Masoud Mohseni, Shun Kanai, Hideo Ohno, Shunsuke Fukami, Luke Theogarajan, Giovanni Finocchio, Supriyo Datta and Kerem Y. Camsari

\thanks{This work was supported in part by a U.S. National Science Foundation grant CCF 2106260, the Office of Naval Research Young Investigator grant and the Semiconductor Research Corporation. S.F. acknowledges JST-CREST Grant No. JPMJCR19K3 and MEXT X-NICS Grant No. JPJ011438. S.K. acknowledges JST-PRESTO Grant No. JPMJPR21B2. A. Grimaldi and G. Finocchio are with the Department of Mathematical and Computer Sciences, Physical Sciences and Earth Sciences, University of Messina, Messina, Italy. M. Mohseni is with Google Quantum AI. S. Kanai, H. Ohno, and S. Fukami are with RIEC, CSIS, WPI-AIMR and Graduate School of Engineering, Tohoku University, Sendai, Japan. S. Datta is with Elmore Family School of Electrical and Computer Engineering, Purdue University, IN, USA. S. Chowdhury, N. A. Aadit, S. Niazi, L. Theogarajan, and K. Y. Camsari are with the Department
of Electrical and Computer Engineering, University of California Santa Barbara, CA, 93106 USA e-mail: (camsari@ucsb.edu).} \vspace{-35pt}}

\maketitle

\begin{abstract}
The transistor celebrated its 75\textsuperscript{th} birthday in 2022. The continued scaling of the transistor defined by Moore's Law continues, albeit at a slower pace. Meanwhile,  computing demands and energy consumption required by modern artificial intelligence (AI) algorithms have skyrocketed. As an alternative to scaling transistors for general-purpose computing, the integration of transistors with unconventional technologies has emerged as a promising path for domain-specific computing. In this article, we provide a full-stack review of probabilistic computing with p-bits as a representative example of the energy-efficient and domain-specific computing movement. We argue that p-bits could be used to build energy-efficient probabilistic systems, tailored for probabilistic algorithms and applications. From hardware, architecture, and algorithmic perspectives, we outline the main applications of probabilistic computers ranging from probabilistic machine learning and AI to combinatorial optimization and quantum simulation. Combining emerging nanodevices with the existing CMOS ecosystem will lead to probabilistic computers with orders of magnitude improvements in energy efficiency and probabilistic sampling, potentially unlocking previously unexplored regimes for powerful probabilistic algorithms. 
\end{abstract}


\IEEEpeerreviewmaketitle

\section{Introduction}

\IEEEPARstart{T}{he} slowing down of the Moore era of electronics  has coincided with the recent revolution in machine learning and AI algorithms. In the absence of steady transistor scaling and energy improvements, training and maintaining large-scale machine learning models in data centers have become a significant energy concern \cite{patterson2021carbon}. The widespread implementation of AI, particularly in industries such as autonomous vehicles \cite{sudhakar2022data}, is an indication that the energy crisis caused by large-scale machine learning models is not just a data center problem, but a global concern. 

Efforts of extending the Moore era of electronics by improving conventional transistor technology continue vigorously. Examples of this approach include 3D heterogeneous integration, 2D materials for transistors and interconnects \cite{chau2019process}, new transistor physics via negative capacitance \cite{wong2018negative,alam2019critical} or entirely new approaches using spintronic and magnetoelectric phenomena to build energy-efficient switches \cite{manipatruni2019scalable,debashis2022low}. 

A complementary approach to extending Moore's Law is to \textit{augment} the existing CMOS ecosystem with emerging, non-silicon nanotechnologies \cite{chen2014emerging,finocchio2021thepromise}. One way to achieve this goal is through heterogeneous CMOS + X architectures where X stands for a CMOS-compatible nanotechnology. For example, X can be magnetic, ferroelectric, memristive or photonic systems. We also discuss an example of this complementary approach, the combination of CMOS with magnetic memory technology, purposefully modified to build probabilistic computers. 

\begin{figure}[!t]
\vspace{0pt}
\centering
\includegraphics[width=0.99\linewidth]{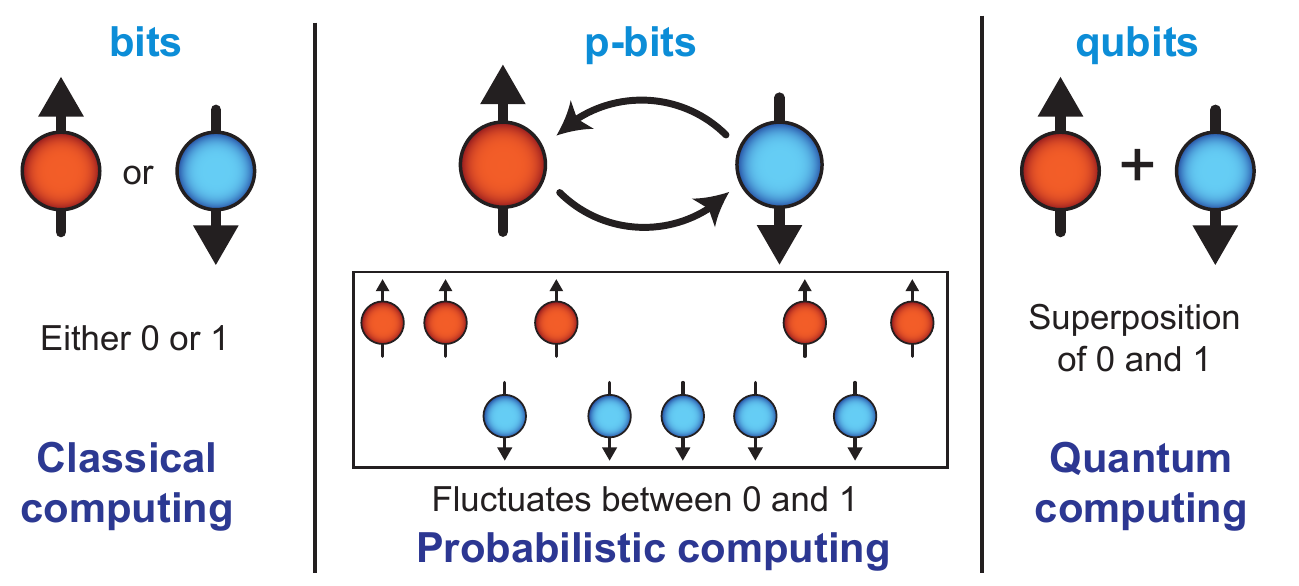}\\
\caption{\textbf{bit, p-bit, and qubit}: Each column shows a schematic representation of the basic computational units of classical computing (left), probabilistic computing (middle), and quantum computing (right). These are, respectively, the bit, the p-bit, and the qubit.}
\label{fig:bpq}
\vspace{-8pt}
\end{figure}

\begin{figure*}[!t]
\vspace{0pt}
\centering
\includegraphics[width=0.99\linewidth]{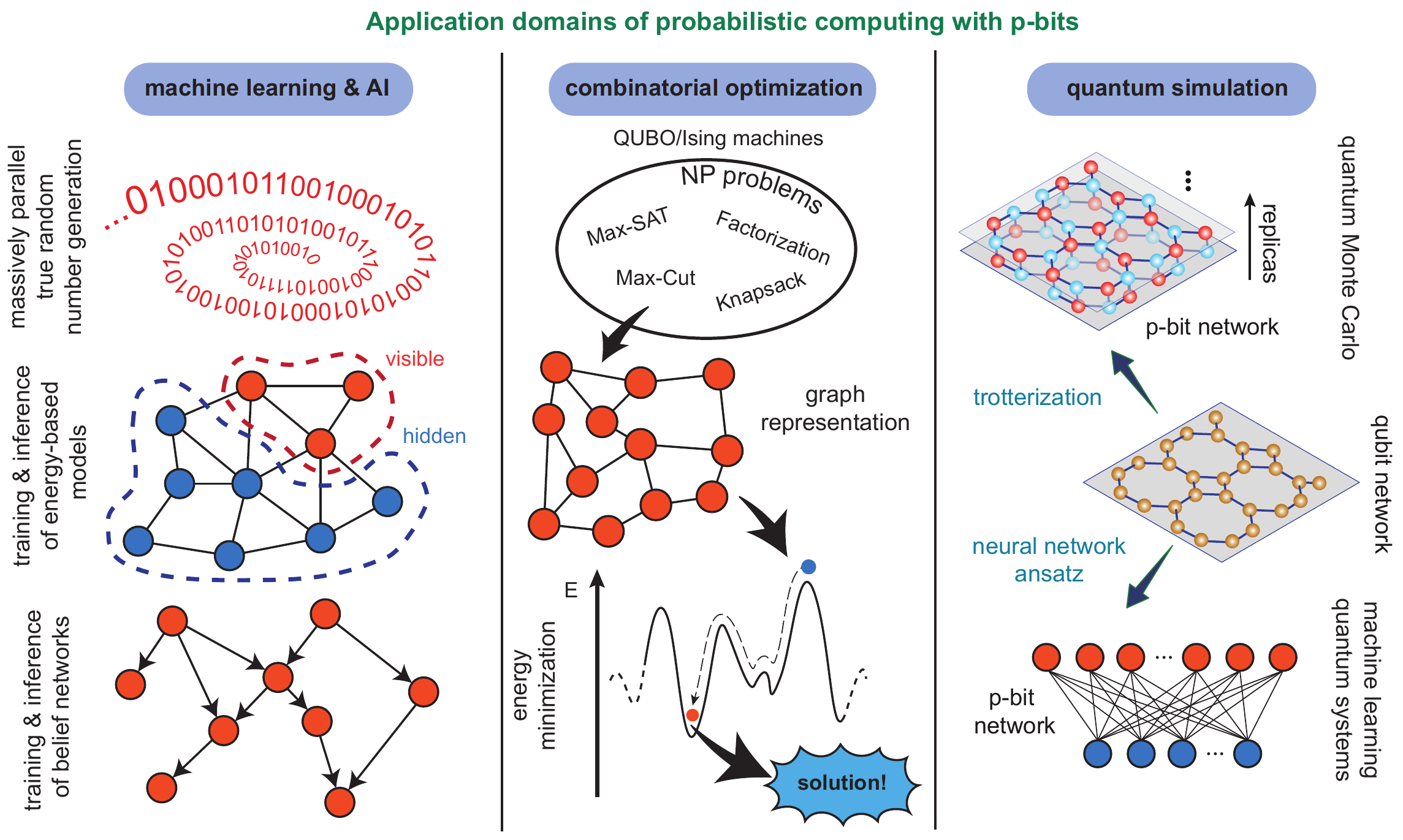}\\
\caption{\textbf{Applications of probabilistic computing}: Potential applications of p-bits are illustrated. The list broadly includes problems in combinatorial optimization, probabilistic machine learning, and quantum simulation.}
\label{fig:app}
\end{figure*}
\section{full-stack view and organization}

Research on probabilistic computing with p-bits originated at the device and physics level, first with stable nanomagnets \cite{behin2016building}, followed by low barrier nanomagnets \cite{sutton2017intrinsic, camsari2017stochasticL}. In \cite{camsari2017stochasticL}, the p-bit was formally defined as a binary stochastic neuron realized in hardware. In both approaches with stable and unstable nanomagnets, the basic idea is to exploit the natural mapping between the intrinsically noisy physics of nanomagnets to the mathematics of general probabilistic algorithms (e.g., Monte Carlo, Markov Chain Monte Carlo). Such a notion of \textit{natural computing} where physics is matched to computation was clearly laid out by Feynman in his celebrated \emph{Simulating Physics with Computers} talk \cite{feynman1982simulating}. Subsequent work on p-bits defined it as an abstraction between bits and qubits (FIG.~\ref{fig:bpq}) with the possibility of different physical implementations. In addition to searching for energy-efficient realizations of single devices, p-bit research has extended to finding efficient architectures (through massive parallelization, sparsification \cite{aadit2022massively} and pipelining \cite{kaiser2021benchmarking}) along with the identification of  promising application domains. This full-stack research program covering hardware, architecture, algorithms, and applications is similar to the related field of quantum computation where a large degree of interdisciplinary expertise is required to move the field forward (see the related reviews Ref.~\cite{misra2022probabilistic,coles2023thermodynamic}). The purpose of this paper is to serve as a consolidated summary of recent developments with new results in hardware, architectures, and algorithms. We provide concrete and previously unpublished examples of machine learning and AI, combinatorial optimization, and quantum simulation with p-bits (FIG.~\ref{fig:app}).   
\section{fundamentals of p-computing}
A large family of problems (FIG.~\ref{fig:app}) can be encoded to coupled p-bits evolving according to the following equations \cite{camsari2017stochasticL}:
\begin{eqnarray}
m_i &=& \mathrm{sign}{\left[\mathrm{tanh}{\left(\beta\,I_i\right)}-r_{[-1,+1]}\right]} \label{eq:pbit}\\
I_i &=& \sum_{j}W_{ij}m_j+h_i \label{eq:synapse}
\end{eqnarray}
where $m_i$ is defined as a bipolar variable ($m \in\{-1,+1\}$), $r$ is a uniform random number drawn from the interval $[-1,1]$, $[W]$ is the coupling matrix between the p-bits, $\beta$ is the inverse temperature and $\{h\}$ is the bias vector. In physical implementations, it is often more convenient to represent p-bits as binary variables, $s_i \in \{0,1\}$. A straightforward conversion of Eq.~(\ref{eq:pbit}) and Eq.~(\ref{eq:synapse}) is possible using  the standard transformation, $m\rightarrow 2s-1$ \cite{borders2019integer}. 

As stated, Eq.~(\ref{eq:pbit}) and Eq.~(\ref{eq:synapse}) do not place any restrictions on the $[W]$, which may be a symmetric or asymmetric matrix. If an update order of p-bits is specified, these equations take the coupled p-bit system to a well-defined steady-state distribution defined by the eigenvector (with eigenvalue $+1$) of the corresponding Markov matrix \cite{camsari2017stochasticL}. Indeed, in case of Bayesian (belief) networks defined by a directed graph, updating the p-bits from parent nodes to child nodes takes the system to a steady-state distribution corresponding to that obtained from the Bayes' Theorem \cite{faria2021hardware}. 

If the $[W]$ matrix is symmetric, one can define an energy, $E$, whose negated partial derivative with respect to p-bit $m_i$ gives rise to Eq.~(\ref{eq:synapse}):
\begin{equation}
E(m_1,m_2, \ldots) = -\left(\sum_{i<j} {W_{ij} m_i m_j} + \sum_i{ h_i m_i}\right) 
\label{eq:energy} 
\end{equation}

In this case, the steady-state distribution of the network is described by \cite{aarts1989simulated}: 
\begin{equation}
p_i = \frac{1}{Z} \exp\left({-\beta E_i}\right)
\end{equation} 
also known as the Boltzmann Law. As such, iterating a network of p-bits described by Eq.~(\ref{eq:pbit}) and Eq.~(\ref{eq:synapse}) eventually approximates the Boltzmann distribution which can be useful for probabilistic sampling and optimization. The approximate sampling avoids the  intractable problem of exactly calculating $Z$. Remarkably, for such undirected networks, the steady-state distribution is invariant with respect to the update order of p-bits, as long as connected p-bits are not updated at the same time (more on this later). This feature is highly reminiscent of natural systems where asynchronous dynamics make parallel updates highly unlikely and the update order does not change the equilibrium distribution. Indeed, this gives the hardware implementation of asynchronous networks of p-bits massive parallelism and flexibility in design. 

The energy functional defined by Eq.~(\ref{eq:energy}) is often the starting point of discussions in the related field of Ising Machines \cite{houshang2020spin,su2022scalable,bhanja2016non,debashis2016experimental,mcmahon2016fully,dutta2021ising,chou2019analog,hitachi2016ising,wang2019_oim,shim2017ising,inagaki2016coherent,Janus_baity2014janus,Hitachi_yamaoka2015,berloff2017realizing,Hitachi_takemoto2019,Toshiba_goto2019combinatorial, Fujitsu_aramon2019physics,mallick2020using,STATICA_yamamoto2020,patel2022logically,afoakwa2020cmos,lu2022scalable,moy20221} with different implementations (see, Ref.~\cite{Mohseni2022} for a comprehensive review). In the case of p-bits, however, we view Eq.~(\ref{eq:pbit}) and Eq.~(\ref{eq:synapse}) more fundamental than Eq.~(\ref{eq:energy}) because the former can also be used to approximate hard inference on \textit{directed} networks while the latter always relies on undirected networks. Compared to undirected networks using Ising Machines, work on directed neural networks for Bayesian inference has been relatively scarce, although there are exciting developments \cite{faria2021hardware,Marsman2015,mccray2020electrically,debashis2020hardware,nasrin2020bayesian,harabi2022memristor,morshed2023deep}. 

Finally, the form of Eq.~(\ref{eq:energy}) restricts the type of interactions between p-bits to a linear one since the energy is quadratic. Even though higher-order interactions ($k$-local) between p-bits are possible \cite{borders2019integer} (also discussed in the context of Ising machines \cite{bybee2022efficient,bashar2022constructing}), such higher-order interactions can always be constructed by combining a standard   probabilistic gate set at the cost of extra p-bits. In our view, in the case of electronic implementation with scalable p-bits,  trading an increased number of p-bits for simplified interconnect complexity is almost always favorable. That being said, algorithmic advantages and the better representative capabilities  of higher-order interactions are  actively being explored \cite{bybee2022efficient,onizawa2021high}.  
\section{hardware: physical implementation of p-bits}
\subsection{p-bit} 
\begin{figure}[!t]
\vspace{0pt}
\centering
\includegraphics[width=\linewidth]{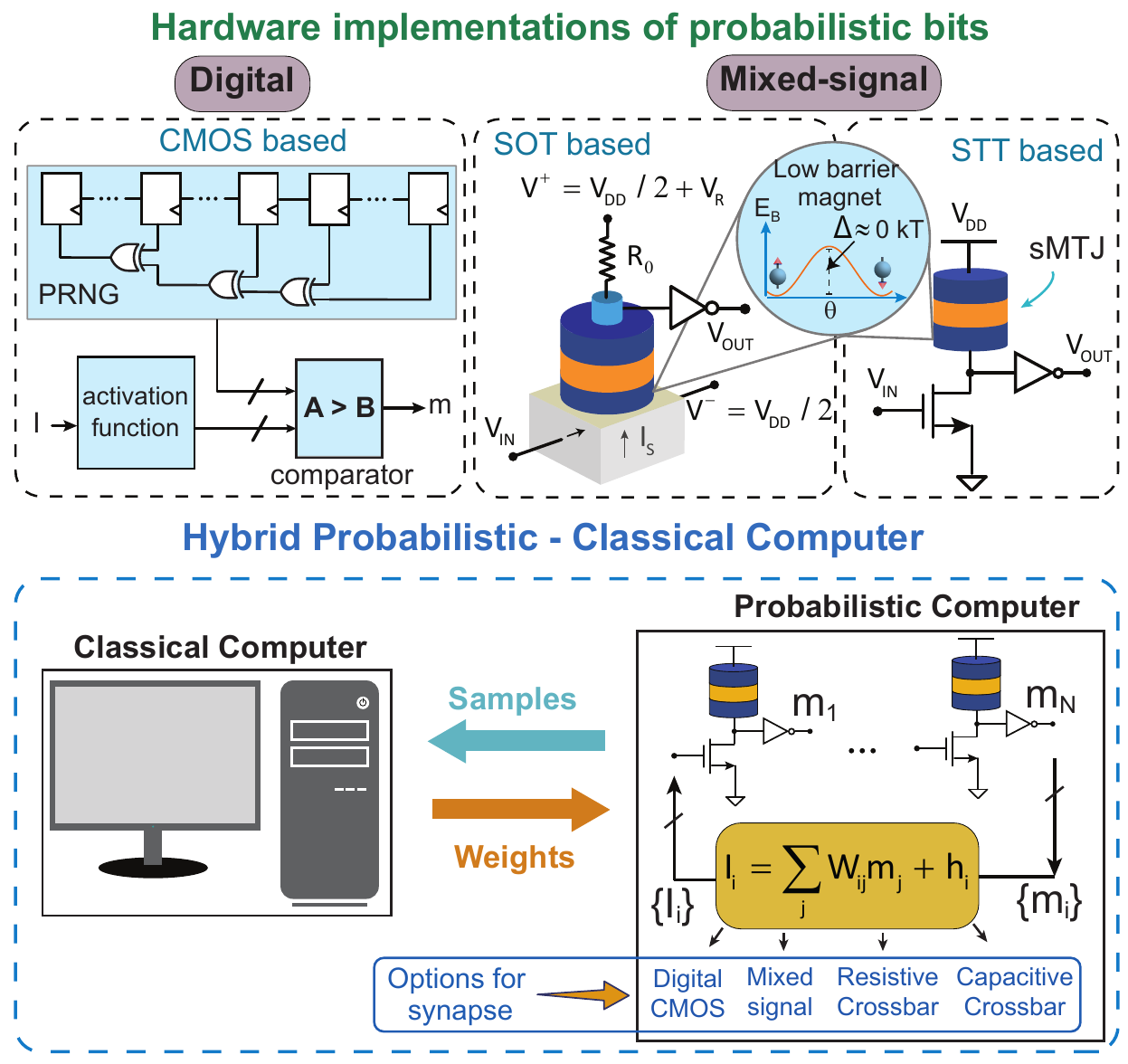}\\
\caption{\textbf{Different hardware options for building a probabilistic computer}: (Top) Various magnetic implementations of a p-bit. These include both digital (CMOS) and mixed-signal implementations (based on, for example, stochastic magnetic tunnel junction with low barrier magnets). (Bottom) A hybrid of classical and probabilistic computing schemes is shown where the classical computer generates weights and programs the probabilistic computer. The probabilistic computer then generates samples accordingly with high throughput and sends them back to the classical computer for further processing. Like the building blocks of p-bits, the synapse of the probabilistic computer can be designed in several ways, including digital, analog, and a mix of both techniques.}
\label{fig:hw}
\vspace{-8pt}
\end{figure}

The p-bit defined in Eq.~(\ref{eq:pbit}) describes a tunable and discrete random number generator. Its physical implementation includes a broad range of options from noisy materials to analog and digital CMOS (FIG.~\ref{fig:hw}). The digital CMOS implementations of p-bits often consist of a pseudorandom number generator (PRNG) ($r$), a lookup table for the activation function (tanh), and a threshold to generate a one-bit output. A digital input with a specified fixed point precision (e.g., 10 bits with 1 sign, 6 integers and 3 fractional) provides tunability through the activation function. Digital p-bits have been very useful in prototyping probabilistic computers up to tens of thousands of p-bits \cite{sutton2020autonomous,kaiser2021probabilistic,aadit2022massively}. They also serve a useful purpose to illustrate \textit{why} analog or mixed-signal implementations of p-bits with nanodevices are necessary. Even using some of the most advanced field programmable gate arrays (FPGA), the  footprint of a digital p-bit is very large: Synthesizing such digital p-bits with pseudo-random number generators (PRNGs) of varying quality of randomness results in tens of thousands of individual transistors.  In single FPGAs that do not use time-division multiplexing of p-bits or off-chip memory, only about 10,000 to 20,000 p-bits with 100,000 weights (sparse graphs with degree 5 to 10) fit, even within high-end devices \cite{aadit2022massively}.

\begin{figure*}[!t]
\vspace{0pt}
\centering
\includegraphics[width=\linewidth]{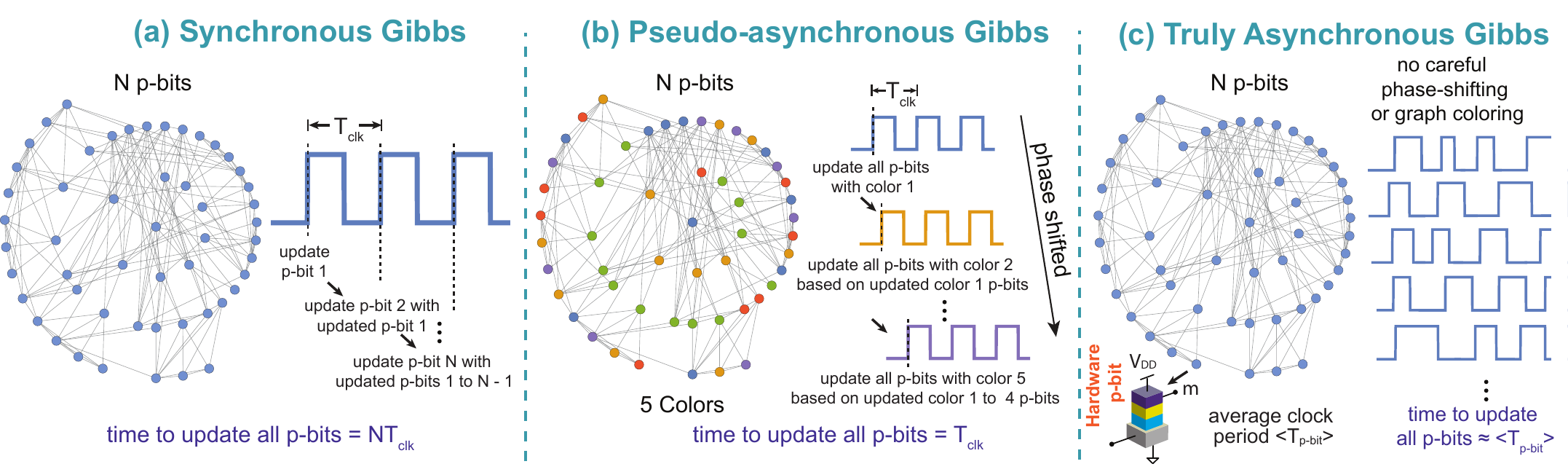}
\caption{\textbf{Architectures of p-computer}: (a)	Synchronous Gibbs: all p-bits are updated sequentially. $N$ p-bits need $N$ clock cycles ($NT_{\rm clk}$) to perform a complete sweep, $T_{\rm clk}$ being the clock period. 
(b)	Pseudo-asynchronous Gibbs: a sparse network can be colored into a few disjoint blocks where connected p-bits are assigned a different color. Phase-shifted clocks update the color blocks one after the other. $N$ p-bits need $\approx$ 1 clock cycle $T_{\rm clk}$ to perform a complete sweep, reducing O($N$) complexity of a sweep to O($1$), where we assume the number of colors $c \ll N$.
(c)	Truly Asynchronous Gibbs: a hardware p-bit (e.g., a stochastic MTJ-based p-bit) provides an asynchronous and random clock with period $\langle T_{\operatorname{p-bit}}\rangle$. $N$ p-bits need approximately 1 clock to perform a complete sweep, as long as synapse time is less than the clock on average. No graph coloring or engineered phase shifting is required.
}
\label{fig:Synch_GIBBS}
\vspace{-6pt}
\end{figure*}

On the other hand, using nanodevices such as CMOS-compatible stochastic magnetic tunnel junctions (sMTJ), millions of p-bits can be accommodated in single cores due to the scalability achieved by the MRAM technology, exceeding 1Gbit MRAM chips \cite{aggarwal2019demonstration,lee20191gbit}. However, before the stable MTJs can be controllably made stochastic, challenges at the material and device level must be addressed \cite{drobitch2019reliability,bandyo2022} with careful magnet design \cite{hassan2021quant,camsari2021double,rahman2022robustness}. Different flavors of magnetic p-bits exist \cite{lv2019experimental,camsari2020charge,chen2022magnetic,rehm2022stochastic}, for a recent review, see Ref.~\cite{Zink2022}. Unlike synchronous or trial-based stochasticity (e.g., \cite{fukushima2014spin}) that requires continuous resetting, the temporal noise of low-barrier nanomagnets makes them ideally suited to build autonomous, physics-inspired probabilistic computers, providing a constant stream of tunably random bits \cite{shao2021implementation}.  
Following earlier theoretical predictions \cite{camsari2017implementing,kaiser2019subnanosecond,hassan2019low}, recent breakthroughs in low-barrier magnets have shown great promise, using stochastic MTJs with in-plane anisotropy where fluctuations can be of the order of nanoseconds  \cite{hayakawa2021nanosecond,safranski2020demonstration,kanai2021}. Such near-zero barrier nanomagnets should be more tolerant to device variations because when the energy-barrier $\Delta$ is low, the usual exponential dependence of fluctuations is much less pronounced. These stochastic MTJs may be used in electrical circuits with a few additional transistors (FIG.~\ref{fig:hw}) to build hardware p-bits. Two flavors of stochastic MTJ-based p-bits were proposed in Ref.~\cite{camsari2017stochasticL} (SOT-based) and in Ref.~\cite{camsari2017implementing} (STT-based). Both of these p-bits have now been experimentally demonstrated in Refs. \cite{borders2019integer,kaiser2022hardware,grimaldi2022experimental} (STT) and in Ref.~\cite{yin2022scalable} (SOT).  While many other implementations of p-bits are possible, from molecular nanomagnets \cite{molpbit} to diffusive memristors \cite{woo2022probabilistic}, RRAM \cite{liu2022probabilistic}, perovskite nickelates \cite{park2022efficient} and others, two additional advantages of the MRAM-based p-bits are the proven manufacturability (up to billion bit densities) and the amplification of room temperature noise. Even with the thermal energy of $kT$ in the environment, magnetic switching causes large resistance fluctuations in MTJs, creating hundreds of millivolts of change in resistive dividers \cite{camsari2017implementing}. Typical noise on resistors (or memristors) is limited by the $\sqrt{kT/C}$ limit which is far lower (millivolts) even at extremely low capacitances ($C$).  This feature of stochastic MTJs ensures that they do not require explicit amplifiers \cite{cheemalavagu2005probabilistic} at each p-bit, which can become prohibitively expensive in terms of area and power consumption. Estimates of sMTJs-based p-bits suggest they can create a random bit using 2 fJ per operation \cite{borders2019integer}.  Recently, a CMOS-compatible single photon avalanche diode-based implementation of p-bits showed similar, amplifier-free operation \cite{whitehead2022cmos} and the search for the most scalable, energy-efficient hardware p-bit using alternative phenomena continues. 

\subsection{Synapse} 
The second central part of the p-computer architecture is the synapse, denoted by Eq.~(\ref{eq:synapse}). Much like the hardware p-bit, there are several different implementations of synapses ranging from digital CMOS, analog/mixed-signal CMOS as well as resistive \cite{10pssyn_xia2016technological} or  capacitors crossbars \cite{li2018capacitor,hassan2019voltage}. The synaptic equation looks like the traditional matrix-vector product (MVP) commonly used in ML models today, however, there is a crucial difference: thanks to the discrete p-bit output (0 or 1), the MVP operation is simply an addition over the active neighbors of a given p-bit. This makes the synaptic operation simpler than continuous multiplication and significantly simplifies digital synapses. In analog implementations, the use of in-memory computing techniques through charge accumulation could be useful with the added simplification of digital outputs of p-bits  \cite{kang2018multi,verma2019memory}. 

It is important to note how the p-bit and the neuron for eventually integrated p-bit applications can be mixed and matched, as an example of creatively combining these pieces, see the FPGA-stochastic MTJ combination reported in Ref.~\cite{grimaldi2022experimental}. The best combination of scalable p-bits and synapses may lead to energy-efficient and large scale p-computers. At this time, various possibilities exist with different technological maturity.

\section{architecture considerations} 

\subsection{Gibbs sampling with p-bits}
\label{sec:Gibbs}
The dynamical evolution of Eqs.~(\ref{eq:pbit}-\ref{eq:synapse}) relies on an \textit{iterated} updating scheme where each p-bit is updated one after the other based on a predefined (or random) update order. This iterative scheme is called Gibbs sampling \cite{geman1984stochastic,koller2009probabilistic}. Virtually all applications discussed in Fig.~\ref{fig:app} benefit from accelerating Gibbs sampling, attesting to its generality. 

In a standard implementation of Gibbs sampling in a synchronous system, p-bits will be updated one by one at every clock cycle as shown in FIG.~\ref{fig:Synch_GIBBS}a. It is crucial to ensure that the effective input each p-bit receives through Eq.~(\ref{eq:synapse}) is computed \textit{before} the p-bit updates. As such, the $T_{\rm clk}$ has to be longer than the time it takes to compute Eq.~(\ref{eq:synapse}). In this setting, a graph with $N$ p-bits will require  $N$ clock cycles ($NT_{\rm clk}$) to perform a complete sweep where $T_{\rm clk}$ is the clock period. This requirement makes Gibbs sampling a fundamentally serial and slow process. 

A much more effective approach is possible by the following observation: even though updates between \textit{connected} p-bits need to be sequential, if two p-bits are not directly connected, updating one of them does not directly change the input of the other through Eq.~(\ref{eq:synapse}). Such p-bits can be updated in parallel without any approximation. Indeed, one motivation of designing \textit{restricted} Boltzmann machines (RBMs, see Ref.~\cite{Hinton2012practical})  over unrestricted BMs  is to exploit this parallelism: RBMs consist of separate layers (bipartite) that can be updated in parallel. However, this idea can be taken further. If the underlying graph is \textit{sparse}, it is often easy to split it into disconnected chunks by coloring the graph using a few colors. Even though finding the \textit{minimum} number of colors is an NP-hard problem \cite{garey1979computers}, heuristic coloring algorithms (such as Dsatur \cite{Brelaz1979}) with polynomial complexity can color the graph very quickly, without necessarily finding a minimum. In this context obtaining the minimum coloring is not critical and sparse graphs typically require a few colors. 

Such an approach was taken on sparse graphs (with no regular structure) to design a massively parallel implementation of Gibbs sampling in Ref.~\cite{aadit2022massively} (FIG.~\ref{fig:Synch_GIBBS}b). Connected p-bits are assigned a different color and unconnected p-bits are assigned the same color. Equally phase-shifted and same-frequency clocks update the p-bits in each color block one by one. In this approach, a graph with $N$ p-bits requires only 1 clock cycle ($T_{\rm clk}$) to perform a complete sweep, reducing O($N$) complexity for a full sweep to O$(1)$, assuming the number of colors is much less than $N$. Therefore, the key advantage of this approach is that the p-computer becomes faster with larger graphs since probabilistic ``flips per second'', a key metric measured by TPU and GPU implementations \cite{FANG20142467,yang2019high} linearly increases with the number of p-bits. It is important to note that these TPU and GPU implementations also solve Ising problems in sparse graphs, however, their graph degrees are usually restricted to 4 or 6, unlike more irregular and higher degree graphs implemented in Ref.~\cite{aadit2022massively}. 

We term this graph-colored architecture the pseudo-asynchronous Gibbs because while it is technically synchronized to out-of-phase clocks, it embodies elements of the truly asynchronous architecture we discuss next. While graph coloring algorithmically increases sampling rates by a factor of $N$, it still requires a careful design of out-of-phase clocks. A much more radical approach is to design a \textit{truly asynchronous} Gibbs sampler as shown in FIG.~\ref{fig:Synch_GIBBS}c. Here, the idea is to have hardware building blocks with naturally asynchronous dynamics, such as a stochastic magnetic tunnel junction-based p-bit. In such a p-bit, there exists a natural ``clock'', $\langle T_{\text{p-bit}}\rangle$, defined by the average lifetime of a Poisson process \cite{debashis2020correlated}. As long as  $\langle T_{\text{p-bit}}\rangle$ is not faster than the average synapse time ($t_{\rm synapse}$) to calculate Eq.~(\ref{eq:synapse}), the network  still updates $N$ spins in a single $\langle T_{\text{p-bit}}\rangle$ timescale. This is because the probability of simultaneous updates is extremely low in a Poisson process and further reduced in highly sparse graphs. 

 In fact, preliminary experiments implementing such truly asynchronous p-bits with ring-oscillator activated clocks show that despite making occasional parallel updates, the asynchronous p-computer performs similarly compared to the pseudo-asynchronous system where incorrect updates are avoided with careful phase-shifting \cite{navid2022nano}. The main appeal of the truly asynchronous Gibbs sampling is the lack of any graph coloring and phase-shift engineering while retaining the same  massive parallelism as $N$ p-bits requires approximately a single $\langle T_{\text{p-bit}}\rangle$ to complete a sweep. Given that the FPGA-based p-computers already provide about a 10X improvement in sampling throughput to optimized TPU and GPUs \cite{aadit2022massively}, such asynchronous systems are promising in terms of scalability. Stochastic MTJ-based p-bits should be able to reach high densities on a single chip. Around 20 W of projected power consumption can be reached considering 20 $\mu$W p-bit/synapse combinations at 1M p-bit density \cite{sutton2020autonomous, bhatti2017spintronics,hassan2021quant}. The ultimate scalability of magnetic p-bits is a significant advantage over alternative approaches based on electronic or photonic devices. 
\begin{figure}[!t]
\vspace{0pt}
\centering
\includegraphics[width=0.95\linewidth]{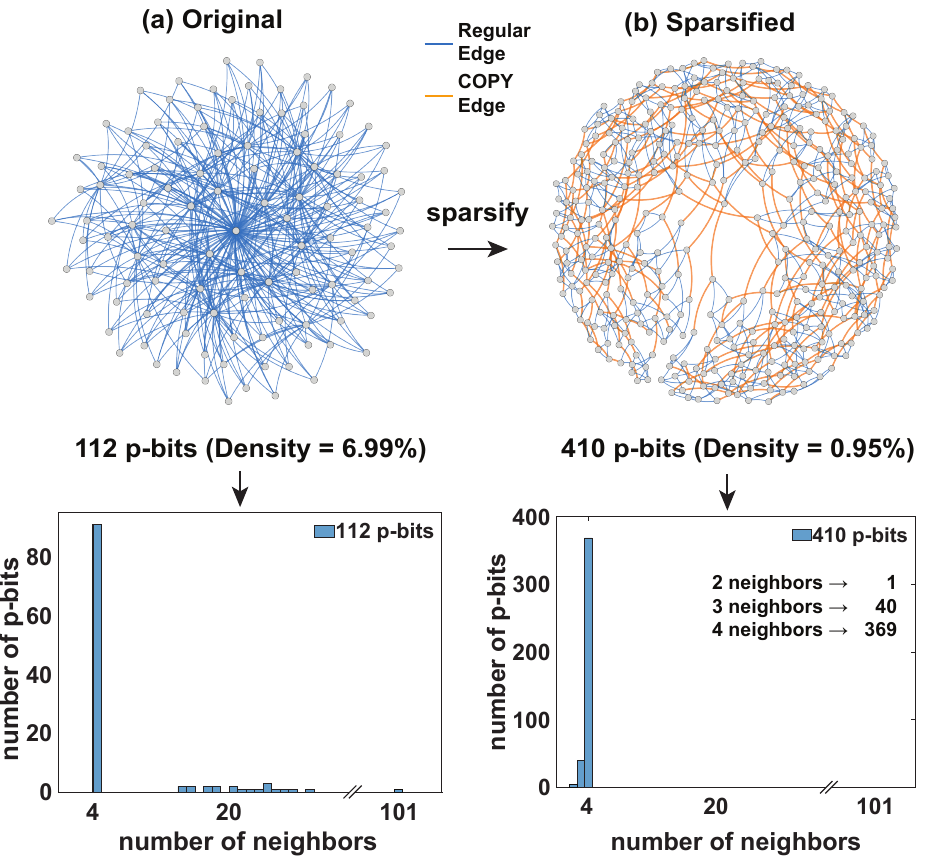}\\
\caption{(a) Original graph of a 3SAT instance uf20-01.cnf \cite{Hoos2000} having 112 p-bits and a graph density of 6.99\%. (Graph density, $\rho = \displaystyle 2|E| /(|V|^2-|V|)$ where $\displaystyle|E|$ = the number of edges and $\displaystyle|V|$  = the number of vertices in the graph). Some of the p-bits have many local neighbors up to 101 neighbors as shown in the histogram which slows down the synapse and the p-bits need to update slowly. (b) Sparsified graph of the same instance having 410 p-bits. COPY gates are inserted between each pair of copies of the same p-bits (COPY edges are highlighted in orange). The graph has a density of $0.95\%$ and the maximum number of neighbors is limited to $4$. The synapse operations are now faster and hence the p-bits can be updated faster. Even though the example shown here starts from a low-density graph, the sparsification algorithm we give is general and applicable to any graph.}
\label{fig:sparsification}
\vspace{-8pt}
\end{figure}
\subsection{Sparsification} 
Both the pseudo-asynchronous and the truly asynchronous parallelism require \textit{sparse} graphs to work well. The first problem is the number of colors: if the graph is dense, it requires a lot of colors,  making the architecture very similar to the standard serial Gibbs sampling. 

The second problem with a dense graph is the synapse time, $t_{\rm synapse}$. If many p-bits have a lot of neighbors, the synapse unit needs to compute a large sum before the next update. If the time between two consecutive updates is $\langle T_{\text{p-bit}}\rangle$, it requires $t_{\rm synapse} \ll \langle T_{\text{p-bit}}\rangle$ to avoid information loss and reach the correct steady-state distribution \cite{pervaiz2017hardware,sutton2020autonomous}. 

However, if the graph is sparse, each p-bit has fewer connections and the updates can be faster without any dropped messages. Any graph can be sparsified using the technique proposed in \cite{aadit2022massively}, similar in spirit to the minor-graph embedding (MGE) approach pioneered by D-Wave \cite{choi2011minor}, even though the objective here is to not find an embedding but to sparsify an existing graph. The key idea is to split p-bits into different copies, using ferromagnetic COPY gates. These p-bits distribute the original connections among them, resulting in identical copies with fewer connections. An important point is that the ground-state of the original graph remains unchanged \cite{aadit2022massively}, so the method does not involve approximations, unlike other sparsification techniques, for example, based on low-rank approximations \cite{sagan2022implementing}.  

FIG.~\ref{fig:sparsification}a shows an example of this process where the original graph of a satisfiability (3SAT) instance has been sparsified as shown in FIG.~\ref{fig:sparsification}b. Irrespective of the input graph size, a sparsified graph has fewer connections locally and thus the neurons hardly ever need to be slowed down. One disadvantage of this technique is the increased  number of p-bits, however, the reduced synapse complexity and the possibility of massive parallelization outweigh the costs incurred by additional p-bits, which we consider to be cheap in scaled, nanodevice-based implementations. 

\begin{figure*}[t]
\vspace{0pt}
\centering
\includegraphics[keepaspectratio,width=0.70\textwidth]{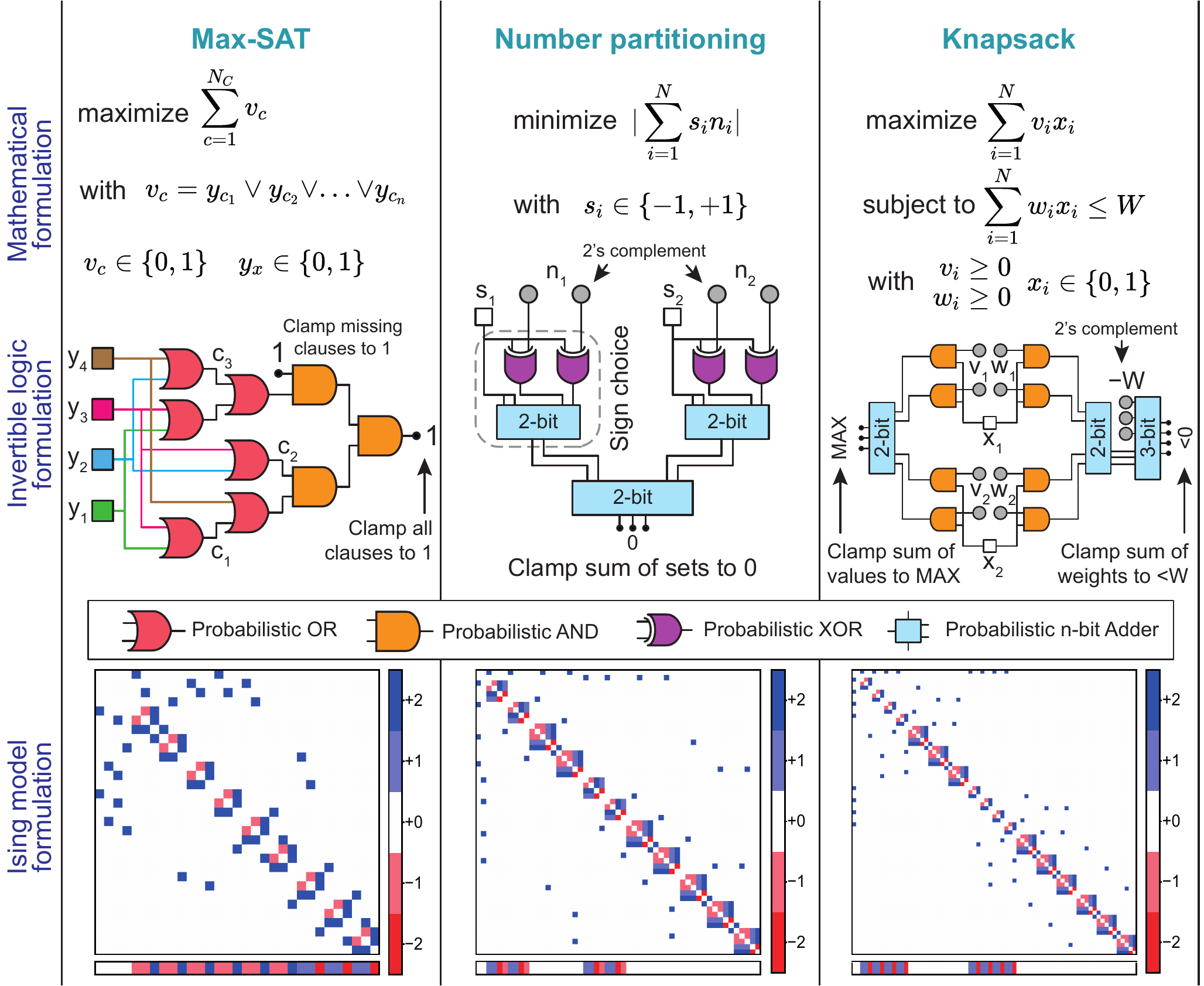}\\
\caption{\textbf{Invertible logic encoding}: The encoding process of three optimization problems, the maximum satisfiability problem (left column), number partitioning (central column), and the knapsack problem (right column), is streamlined and visually summarized into three steps: a problem first has to be condensed into a concise mathematical formulation (upper part of the image); then, an invertible Boolean circuit that topologically maps the problem is conceived; finally, the invertible Boolean circuit is converted into an Ising model using probabilistic AND/OR/NOT gates \cite{aadit2022massively}.}
\label{fig:figINV}
\vspace{-6pt}
\end{figure*}

\section{Algorithms and Applications}

\subsection{Combinatorial Optimization via Invertible Logic}

When using the Ising model to solve an optimization problem, the first step is to provide a mapping between the Ising model and the problem to be solved. Early work on quantum annealing stimulated by D-Wave's quantum annealers generated a significant amount of useful research in this area \cite{lucas2014ising}, some of which are being adopted by quantum-inspired classical annealers. There are usually many different ways to find a mapping, for example, some strategies may employ more nodes than others to encode the same instance, while others might result in graphs with topology unsuited to the computational architecture of choice. In this context, the invertible logic approach introduced in  Ref. \cite{camsari2017stochasticL} stands out for its flexibility and sparse encodings.

The process of mapping an instance into an Ising model can be broadly summarized into three steps, as illustrated by FIG.~\ref{fig:figINV}. In the figure, the steps of the invertible logic encoding of three combinatorial optimization problems, maximum satisfiability (left column), number partitioning (middle column), and knapsack (right column), are shown. First, each problem is formalized into a tight mathematical formulation (top row). Next, the problem is mapped into an invertible Boolean logic circuit (central row), meaning that each logic gate can be operated using any terminals as input/output nodes (similar to those discussed in the context of quantum annealing \cite{andriyash2016boosting,biamonte2008nonperturbative} and memcomputing \cite{traversa2017polynomial,di2022memcomputing}). Finally, the probabilistic circuit is algorithmically encoded into an Ising model (bottom row). Each logic gate has several Ising encodings that map the energy landscape of its logic operator. After the Boolean logic formulation of a problem, this step can be automated in standardized synthesis tools. The overall approach results in relatively sparse circuits, as illustrated in the bottom row of FIG.~\ref{fig:figINV} where all three problems show similarly sparse matrices [W], with bias vectors $\{h\}$ shown under.

The key advantage of this approach, compared to heuristic and dense formulations of Ref.~\cite{lucas2014ising}, is due to the generality of Boolean logic, quite similar to how present-day digital VLSI circuits are constructed in sparse, hardware-aware networks using billions of transistors. As such, much of the existing ecosystem of high-level synthesis can be directly used to find invertible logic-based encodings for general optimization problems. 

\begin{figure}[!t]
\vspace{0pt}
\centering
\includegraphics[keepaspectratio,width=0.77\columnwidth]{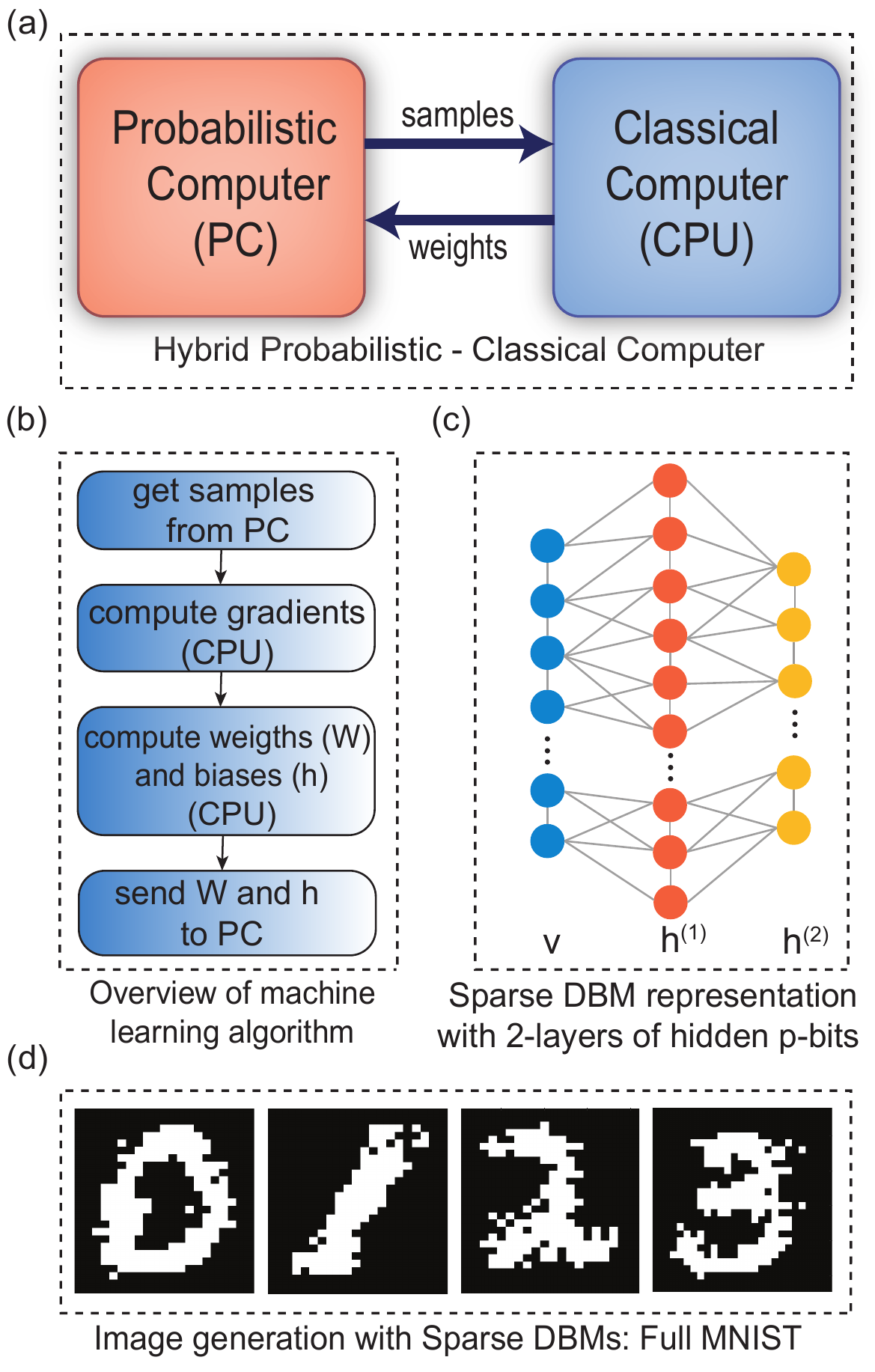}\\
\caption{\textbf{Generative neural networks with p-bits}: (a) Hybrid computing scheme with probabilistic computer and classical computer is demonstrated where the probabilistic computer generates samples according to the weights given by CPU with a sampling speed of around 100 flips/ns. (b) shows the overview of the learning procedure for the hybrid set-up. Receiving the samples from the probabilistic computer, the CPU computes the gradient, updates the weights and biases, and sends them back to the probabilistic computer until converged. (c) Sparse Deep Boltzmann Machine (DBM) is utilized here as a hardware-aware graph that can be represented with  multiple hidden layers of p-bits. Both inter-layer and intra-layer connections are allowed between visible and hidden units. (d) The images shown here are generated with a sparse DBM of 4264 p-bits after training the network with full MNIST. The label p-bits are clamped to a specific image and the network evolves to that image by annealing the system from $\beta=0$ to $\beta=5$ with a step size of 0.125.} \vspace{-4pt} \label{fig:ML_fig}\end{figure} 

\subsection{Machine Learning: energy-based models}

Energy-efficient machine learning (ML) with Boltzmann Machines (BMs) is a promising application for probabilistic computers with a recent experimental demonstration in Ref.~\cite{kaiser2022hardware}. Mainstream ML algorithms are designed and chosen with CPU implementation in mind and hence some models are heavily preferred over others even though they are often less powerful. For example, the use of RBMs over the more powerful unrestricted or deep BMs is motivated by the former's efficient software implementation in synchronous systems. However, by exploiting the technique of sparsity and massively parallel architecture described earlier, fast Gibbs sampling with DBMs can dramatically improve the state-of-the-art machine learning applications like visual object recognition and generation, speech recognition, autonomous driving, and many more \cite{salakhutdinov2009deep}. Here we present an example where a sparse DBM is trained with MNIST handwritten digits (FIG.~\ref{fig:ML_fig}). We randomly distribute the visible and hidden units on the sparse DBM with massively parallel pseudo-asynchronous architecture which yields multiple hidden layers as shown in FIG.~\ref{fig:ML_fig}(c). 

Contrasting with earlier unconventional computing approaches where the MNIST dataset is reduced to much smaller sizes \cite{adachi2015application,manukian2019accelerating}, we show how the full MNIST dataset (60,000 images and no down-sampling)  can be trained using p-computers in FPGAs. We use 1200 mini-batches having 50 images in each batch to train the network using the contrastive divergence (CD) algorithm. The process of learning is accomplished using a hybrid probabilistic and classical computer setup. The classical computer computes the gradients and generates new weights while the p-computer generates samples according to those weights (FIG.~\ref{fig:ML_fig}(b)). During the positive phase of sampling, the p-computer operates in its clamped condition under the direct influence of the training samples. In the negative phase, the p-computer is allowed to run freely without any environmental input. After training, the deep network not only can classify images but also generate images. For any given label, the network can create a new sample (not present in the training set) (FIG.~\ref{fig:ML_fig}(d)). This is an important feature of energy-based models and is commonly demonstrated with diffusion models \cite{sohl2015deep}.

\begin{figure*}[!t]
\vspace{0pt}
\centering
\includegraphics[keepaspectratio,width=0.9\textwidth]{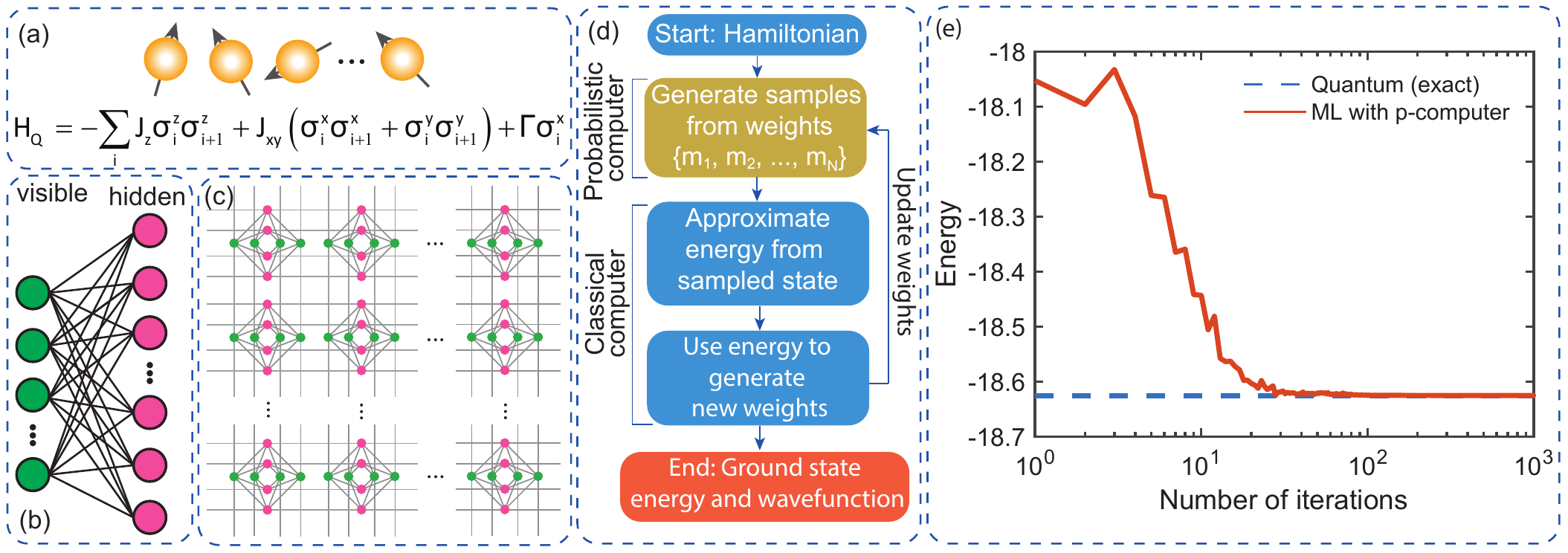}\\
\caption{\textbf{Machine learning quantum systems with p-bits}:  (a) Heisenberg Hamiltonian with a transverse field ($\Gamma = +1$) is applied to a FM coupled ($J_{Z} = +1$ and $J_{xy} = +0.5$) linear chain of twelve ($12$) qubits with periodic boundary.  (b) To obtain the ground state of this quantum system, a restricted Boltzmann machine is employed with 12 visible and 48 hidden nodes, where all nodes in the visible layer are connected to all nodes in the hidden layer. (c) This machine learning model is then embedded onto a hardware amenable sparse p-bit network arranged in a chimera graph  using minor graph embedding. We use a coupling strength of 1.0 among the replicated visible and hidden nodes in the embedded p-bit network. (d) An overview of the machine learning algorithm and the division of workload between the probabilistic and classical computers in a hybrid setting is shown. (e) The FPGA emulation of this probabilistic computer performs variational machine learning in tandem with a classical computer, converging to the quantum (exact) result as shown.}\label{fig:figQML}
\vspace{-4pt}
\end{figure*}

\subsection{Quantum simulation}
One primary motivation for building quantum computers is to simulate large quantum many-body systems and understand the exotic physics offered by them \cite{ma2020quantum}. Two major challenges with 
quantum computers are the necessity of using cryogenic operating temperatures and the vulnerability to noise, rendering quantum computers impractical, especially considering practical overheads \cite{babbush2021focus}. Simulating these systems with classical computers is often extremely time-consuming and mostly limited to  small systems. One potential application of p-bits is to provide a room-temperature solution to boost the simulation speed and potentially enable the simulation of large-scale quantum systems. Significant progress has been made toward this end in recent years.

\subsubsection{Simulating quantum systems with Trotterization}
\noindent One approach is to build a p-computer enabling the scalable simulation of sign-problem-free quantum systems by accelerating standard Quantum Monte Carlo (QMC) techniques \cite{camsari2019scalable}. The basic idea is to replace the qubits in the original lattice with hardware p-bits and replicate the new lattice according to the Suzuki-Trotter transformation \cite{suzuki1976relationship}. Recently, the convergence time of a 2D square-octagonal qubit lattice initially prepared in a topologically obstructed state was compared among a CPU, a physical quantum annealer \cite{king2021scaling} and a p-computer (both digital and analog) \cite{chowdhury2022accelerated}. For this particular problem, it was shown that an FPGA-based p-computer emulator can be around 1000 times faster than an optimized C++ (CPU) program. Based on SPICE simulations of a small p-computer,
we project that significant further acceleration should be
possible with a truly asynchronous implementation. Probabilistic computers can be used for quantum Hamiltonians beyond the usual Transverse Field Ising Model, such as the antiferromagnetic Heisenberg Hamiltonian \cite{chowdhury2019probabilistic} and even for the emulation of gate-based quantum computers \cite{chowdhury2020emulating}. However, for generic Hamiltonians, (e.g.,  random circuit sampling), the number of samples required in naive implementations seem to grow exponentially \cite{chowdhury2020emulating} due to the notorious \emph{sign-problem} \cite{troyer2005computational}. However, clever basis transformations \cite{aharonov2022polynomial}
might mitigate or cure the sign-problem \cite{hangleiter2020easing} in the future.

\subsubsection{Machine learning quantum many-body systems}
\noindent  With the great success of machine learning and AI algorithms, training stochastic neural networks (such as Boltzmann machines) to approximately solve the quantum many-body problem starting from a variational guess  has generated great excitement \cite{carleo2017solving,Cai2018,saito2017} and is considered to be a fruitful combination of quantum physics and machine learning \cite{sarma2019machine}. These algorithms are typically implemented in high-level software programs, allowing users to choose from various network models and sizes according to their needs. However, as with classical machine learning, the difficulty of training strongly hinders the use of deeper and more general models.  With scaled p-computers using millions of magnetic p-bits, massively parallel and energy-efficient \textit{hardware} implementations of the more general unrestricted/deep BMs may become feasible, paving the way to simulate practical quantum systems. 

To demonstrate one such example of this approach, we show how p-bits laid out in sparse, hardware-aware graphs can be used for machine learning quantum systems (FIG.~\ref{fig:figQML}). The objective of this problem is to find the ground state of a many-body quantum system, in this case, a 1D FM Heisenberg Hamiltonian with an external transverse field.  We start with an RBM, which is one of the simplest neural network models, and use its functional form as the variational guess for the ground state probabilities (the wavefunction is obtained by taking the square root of probabilities according to the Born rule). A combination of probabilistic sampling and weight updates gradually adjusts the variational guess such that the final guess points to the ground state of the quantum Hamiltonian. Emulating this variational ML approach with p-bits requires a few more steps. An RBM network contains all-to-all connections between the visible and hidden layers which is not conducive for scalable p-computers because of the large fanout demanded by the all-to-all connectivity. An alternative is to map the RBM onto a sparse graph through minor graph embedding  \cite{choi2011minor}. Using a hybrid setup with fast sampling in a probabilistic computer coupled with a classical computer, the iterative process of sampling and weight updating can then be performed. The key advantage of having a massively parallel and fast sampler is the selection of higher-quality states of the wave function to update the variational guess. FIG.~\ref{fig:figQML} shows an example simulation of how a p-computer learns the ground state of a 1D FM Heisenberg model. The scaling of p-computers using magnetic p-bits may allow much larger implementations of quantum systems in the future. 

\subsection{Outlook: algorithms and applications beyond }

Despite the large range of applications we discussed in the context of p-bits, much of the sampling algorithms have been either standard MCMC or generic simulated annealing-based approaches. Future possibilities involve more sophisticated sampling and annealing algorithms such as parallel tempering (PT) (see Ref.~\cite{aadit2021computing,grimaldi2022spintronics}
for some initial investigations). Further improvements to hardware implementation include  \textit{adaptive} versions of PT \cite{desjardins2010adaptive} as well as sophisticated nonequilibrium Monte Carlo (NMC) algorithms \cite{mohseni2021nonequilibrium}. Ideas involving \textit{overclocking} p-bits such that they violate the $t_{\rm synapse} \ll \langle T_{\text{p-bit}}\rangle$ requirement for further improvement \cite{aadit2022massively} or sharing synaptic operations between p-bits \cite{park2022efficient} could  also be useful. A combination of these ideas with \textit{algorithm-architecture-device} co-design may lead to orders of magnitude improvement in sampling speeds and quality. In this context, as a sampling throughput metric, increasing flips/ns is an important goal. In addition, solution quality, the possibility of cluster updates or algorithmic techniques also need to be considered carefully. Given the plethora of approaches from multiple communities, we also hope that model problems, benchmarking studies comparing different Ising machines, probabilistic accelerators, physical annealers and dynamical solvers will be performed in the near future from all practitioners, including ourselves. 

We believe that the co-design of algorithms, architectures, and devices for probabilistic computing may not only help mitigate the looming energy crisis of machine learning and AI but also lead to systems which may unlock previously inaccessible regimes using powerful probabilistic (randomized) algorithms \cite{buluc2021randomized}. Just as the emergence of powerful GPUs made the well-known backpropagation algorithm flourish, probabilistic computers could lead us to the previously unknown territory of energy-based AI models, combinatorial optimization and  quantum simulation. This research program requires a concerted effort and interdisciplinary expertise from all across the stack and ties into the larger vision of unconventional computing forming in the community \cite{finocchio2023roadmap}. 

\ifCLASSOPTIONcaptionsoff
  \newpage
\fi

\clearpage 
\balance{
\bibliographystyle{IEEEtran}}

% Generated by IEEEtran.bst, version: 1.14 (2015/08/26)

\end{document}